# Fabrication and micro-Raman spectroscopy of arrays of copper phthalocyanine molecular-magnet microdisks

Jiří Liška,*,[†] Tomáš Krajňák,[†,‡] Peter Kepič,[†,‡] Martin Konečný, [§,†,‡] , Martin Hrtoň, [†,‡] Vlastimil Křápek, [†,‡], Zdeněk Nováček, [†,‡]  Lorenzo Tesi, [§] Joris van Slageren, [§] Jan Čechal, [†,‡] Tomáš Šikola,[†,‡]

[†]   *Central European Institute of Technology, Brno University of Technology, Purkyňova 123, 612 00, Brno, Czech Republic*

[‡]   *Institute of Physical Engineering, Faculty of Mechanical Engineering, Brno University of Technology, Technická 2, 616 69, Brno, Czech Republic*

[§]   *Institut für Physikalische Chemie, Universität Stuttgart, Pfaffenwaldring 55, D-70569 Stuttgart, Germany*

**Corresponding Author**      *E-mail: jiri.liska@ceitec.vutbr.cz (J.L.).

# ABSTRACT

Phthalocyanines as organic semiconductors and molecular magnets provide plenty of industrial or high-tech applications from dyes and pigments up to gas sensors, molecular electronics, spintronics and quantum computing. Copper phthalocyanine (CuPc) belongs among the most used phthalocyanines, typically in the form of powder or films but self-grown nanowires are also known. Here we describe an opposite, i.e., top-down approach based on fabrication of ordered arrays of CuPc microstructures (microdisks) using electron beam lithography and other steps. Among critical points of this approach belongs a choice of a proper resist and a solvent. Fabricated CuPc microdisks have a diameter of 5 μm and heights from 7 up to 70 nm. Micro-Raman spectroscopy of the films and microdisks reveals a crystalline β phase associated with a paramagnetic form. Additional measurements with an increasing laser power show a significant shift ($\Delta\omega \sim 7.1$ cm$^{-1}$) and broadening of a peak at 1532 rel. cm$^{-1}$ corresponding to the phonon $B_{1g}$ mode. The observed smooth changes exclude a phase transition and confirm the thermally stable polymorph. Our versatile fabrication technique using the common lithographic resist brings new possibilities for the fabrication of various micro/nanostructures such as micromagnets, heterostructures or organic electronic devices.

# 1. INTRODUCTION

Phthalocyanines are a large group of materials with favourable optical, chemical, electronic, and magnetic properties. They belong to organic semiconductors [1,2] and molecular magnets [3,4]. Copper (Cu) phthalocyanine (CuPc) is a well-known example of transition-metal phthalocyanines (TMPc, TM = Co, Cu, Fe, Mn, Ni, V, Zn). CuPc, similarly as other TMPc, occurs typically in α and β crystalline phase, although, several other polymorphs, such as ε phase [5] have been proposed for CuPc. The α and β phases differ in their crystallographic properties, e.g. the angle between the planar molecules and the b-axis is 65° and 45° for α and β phase, respectively [3,6] and also, e.g., in thermal stability or magnetic ordering. While the metastable α phase of CuPc shows antiferromagnetic behaviour, the thermally stable β phase shows paramagnetic properties [3]. However, even ferromagnetic ordering was observed in the α phase of CuPc with surprisingly higher magnetization at 280 K than at 5 K [4].

Since the first synthesis in the early 20th century, phthalocyanines have been used as dyes and pigments [7]. In this field, the CuPc substance had a significant role with its characteristic blue colour tones or even green colour tones for halogenated CuPc. However, nowadays phthalocyanines are also considered as promising candidates for other applications. Namely CuPc in its various forms were used in e.g. CD/RW (ReWritable) disks and photoconductors in Laser Printers [8]. CuPc have highly promising electronic properties for organic-thin-film field-effect transistors OFETs [9,10] and organic solar cells [11]. Their chemical properties are utilized among others in form of gas sensors, e.g. $NO_2$ [12–15] and $CCl_4$ [16], or $NH_3$ [17]. Other applications were proposed in spintronics and quantum computing [18].

For industrial applications as dye and colorant, the production of chemically synthesized CuPc in the form of powder is sufficient and meets the goal of high gain and low price [7]. However, in the last two-three decades, there has been an increased interest in a tailored preparation of CuPc thin films and micro/nanostructures, especially to fulfil the requirements for organic semiconductor-based photovoltaics, spintronics and molecular electronics applications. It was shown, that even by chemical synthesis the CuPc can be obtained in the form of self-assembled nanowires, nanorods or nanotubes [19–24]. Similar structures can be also prepared by temperature gradient sublimation [25], which is furthermore a suitable technique for single crystals formation [3,26,27]. Various CuPc derivates are also used in form of powder for thin film preparation. Despite the fact that several techniques of CuPc film deposition were tested, such as a pulsed laser deposition (PLD) [28] or a glow-discharge-induced sublimation (GDS) [29], the most traditional way remains a thermal evaporation (or better to call sublimation) of CuPc powder in vacuum [6,10,28,30–32] or in ultrahigh vacuum known as an organic molecular beam deposition (OMBD) [3]. These works showed that a purity of used powder, deposition temperature, base pressure, choice of substrate and its surface properties or substrate temperature during or after deposition have a significant influence on the form and quality of the final films. Further variations of this evaporation technique can be achieved in the case of vacuum evaporation under magnetic [33] or electric field [34] and even the use of microgravity conditions in space was tested for CuPc growth [35]. Multilayer thin CuPc films doped by hexacyano-trimethylene-cyclopropane (CN6-CP) [36] or chloro-substituted copper phthalocyanines $CuPcCl_x$ (where x is equal 4, 8 or 16) [37] are examples of more complex film preparation.

The thin film with metastable α phase of CuPc can be prepared by thermal evaporation on substrates held close to the room temperature and with the deposition pressure below $10^3$ Pa [30,38]. On the contrary, the thermally stable β phase requires substrate temperature above ~210°C or higher pressure. Post-annealing of CuPc films has an impact on their morphology [39] and can drive irreversible phase transition between α and β phase, as was observed several times. Reported phase transition temperatures have a quite broad distribution centred around 250°C (523 K). The process of transition should be kinetically activated which means combination of temperature, duration and CuPc volume. The activation is easier for smaller crystals. The lower limit is 210°C (483 K), which is the substrate temperature during deposition to grow films directly in the β phase [38]. The transition was observed e.g. at 240°C (513 K) but the films were transformed only partially [40]. With increasing the crystallite size, the required temperature grows up to the bulk value of 300°C (573 K) [41]. This is confirmed by observed transition temperatures at 250°C (523 K) and 290°C (563 K) for films with small and large grain-sizes [29]. CuPc films were also annealed at a higher temperature of 320°C (593 K) [3], or even under the $N_2$ flux at 350 °C (623 K) (the maximum found value) [30,42]. CuPc nanorods are a special case, in which the transition temperature is lowered up to 183°C (456 K) [19]. To complete driving forces, the phase transition from α to β form can be reached using treatment of some aromatic hydrocarbons like a xylene, more references in [43].

Structural properties of CuPc are frequently studied using X-ray diffraction, which allows to distinguish polymorphic phases and phase transition [3,29,30]. Identification of phases can be achieved also using Raman spectroscopy [5,19] based on observation of peaks from several vibrational modes in the wavenumber range 700-1600 cm$^{-1}$. Complementary information is provided by infrared absorption spectroscopy measured by Fourier transform infrared spectroscopy (FTIR) [6,31]. UV-vis-NIR spectroscopy shows that CuPc contains two main absorption bands related to electron transitions, B-band and Q-band at around 335 and 650 nm, respectively [31,33]. X-Ray photoelectron spectroscopy (XPS) combined with an ultraviolet photoemission spectroscopy (UPS) allows to determine chemical bonds and element composition, together with the electronic properties and binding energies of CuPc [42,44]. Theoretical studies of electronic or magnetic properties of CuPc are presented in Refs. [45–47].

The above-mentioned applications utilize CuPc in form of powders, (thin) films or self-assembled micro/nanostructures (nanowires, nanorods or nanotubes). However, for many future devices, a precise positioning and shape engineering of molecular structures will be required, which in turn brings about a need for a controlled micro/nanofabrication technique for CuPc nanostructures. Our aim is to fabricate a device composed of CuPc microdisks serving as the best candidate sample to test a novel THz tip-enhanced electron paramagnetic resonance setup [48]. Organic semiconductors and molecular magnets frequently interact with common solutions used during microfabrication techniques; therefore, only a limited number of articles describe the successful microfabrication process. A shadow mask deposition that provides molecule magnet microstructures with sizes above 1.7 µm was reported [49]. In this case, a copper TEM grid was utilized to form a dot microarray from terbium(III) bis-phthalocyaninato neutral complex. Another approach that combined photolithography, deposition and lift-off was enough to reach a resolution down to 4 µm for Poly(3,4-ethylenedioxythiophene):poly(styrene sulfonate) (PEDOT:PSS) [50], detailed discussion on this topic and other references therein. To complete the list by CuPc material, Aoki et al. mentioned fabrication of spatially limited CuPc film using a square metal mask with an opening of 30×30 µm$^2$, prepared before the CuPc deposition [10]. Here, on the other hand, we propose a fabrication approach based on the lithographic patterning (electron beam lithography) that can be used as versatile recipe for the production of a variety of CuPc micro and nanostructures and can be easily combined with common fabrication techniques of metal or dielectric films or structures.

## 2. MATERIAL AND METHODS

**2.1 Fabrication of samples**. Our approach for fabrication of CuPc microdisks involves resist spin-coating, electron beam lithography (EBL), resist development, CuPc deposition, and lift-off (see schematics in Fig. 1). The fabrication was realised with the equipment available in the CEITEC Nano RI cleanrooms with ISO class 5 or 8.

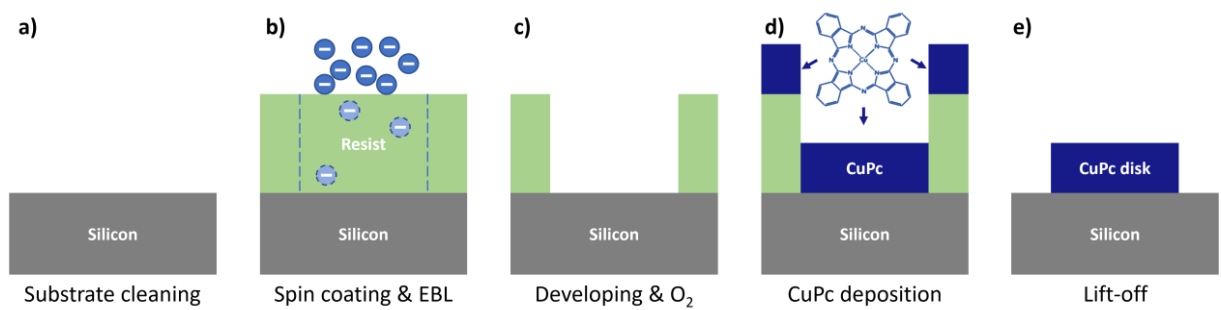

**Fig. 1.** Fabrication process of CuPc microdisks illustrated for a silicon substrate.

At the beginning, single silicon crystal Si(100) or fused silica double-side polished wafers were cut by Laser Dicer (Oxford Lasers Ltd.) to get 3×3×0.5 or 10×10×0.5 mm$^3$ substrates. Prior the resist coating, substrates were cleaned in an acetone and isopropyl alcohol ultrasonic bath, dried by N$_2$ flow, and baked on a hot plate at 150°C (Fig. 1a). A thin film of the positive tone electron beam polymethyl methacrylate (PMMA) resist (AR-P 649.04 or 679.04, Allresist GmbH) was spin-coated on top of the clean substrate surface and baked out at 150°C on a hotplate for 3 min with a final thickness of 190 and 310 nm, respectively. EBL was performed by a scanning electron microscope SEM Mira3 (Tescan Orsay Holding) equipped with a laser interferometer stage (Raith GmbH). The 3×3 mm$^2$ area of the resist surface was patterned by an array of circles with a diameter of 5 μm and pitch of 10 μm using the 30 keV electron beam and set area dose of 200 μC/cm$^2$ (Fig. 1b). A high probe current of 4.7 nA and large writing field (500×500 μm$^2$) were selected to reach a fast patterning (30 min/9 mm$^2$). In the case of a non-conductive fused silica substrate, the conductive resist film Electra 92 (AR-PC 5090.02) was spin-coated on the PMMA film to prevent charging and then the sample was baked out at 110°C for 2 min before EBL. After EBL, the Electra film was removed using demineralized water. The developing procedure involved bath in the developer AR 600-56 (methyl isobutylketone, for 3 min), in the stopper AR 600-60 (isopropylalcohol, for 30 s), and flush in demineralized water (for 30 s). This resulted in the creation of cylindrical microwells in the PMMA resist (Fig. 1c). Before subsequent CuPc deposition, O$_2$ plasma treatment (PlasmaPro NGP 80, Oxford Instruments Ltd.) with set parameters: power of 50 W, O$_2$ flow of 20 sccm, working pressure of 40 mTorr, DC Bias of 258 V, and determined PMMA etching rate of 2.1 nm/s, was applied for tens of seconds as a resist stripper, which helped to locally improve the adhesion between the CuPc and the substrate surface in the microwells.

Such pre-patterned samples were fixed to a sample holder and inserted into a UHV-deposition chamber which is part of a large complex UHV system. A crystalline CuPc powder (Copper(II) phthalocyanine, CAS No. 147-14-8, Alfa Aesar) was degassed and heated in the crucible of an effusion cell to 440-460°C. During the CuPc film deposition, the samples were kept at room temperature, under the pressure in the range of $10^{-6}$-$10^{-7}$ Pa, and CuPc atoms sublimated from the crucible were deposited on them (Fig. 1d). In the first deposition run, tens-of-minutes-long depositions led to CuPc films with a thickness up to 20 nm (average deposition rates were ~ 30 nm/h). The second run with lower average deposition rates 1.0-1.7 nm/h and longer deposition times lasting for tens of hours resulted in thicker CuPc films with a thickness of 70 nm. The deposition rate depends on the amount and distribution of the evaporated material in the crucible, as well as the crucible temperature during deposition and degassing steps.

After the deposition, the PMMA resist together with CuPc layer on top, was removed during the lift-off process using the acetone bath for units or tens of minutes (Fig. 1e). To avoid an expected degradation of CuPc microstructures, the lift-off process duration was kept as short as possible. It has been found out that for a complete removal of the bare PMMA film (i.e. without CuPc) it is necessary to let the sample in acetone for at least 30 second. Successful lift-off of a 7 nm-thick CuPc film on the PMMA microwells takes at least 1 minute and the acetone bath must be more prolonged for thicker CuPc films on the patterned PMMA. No significant erosion of the CuPc structures/film was anyway observed even after a tens-of-minutes-long acetone bath. On the other hand, after applying a seconds-long ultrasound treatment, which is very often used to release structures during the lift-off process, a lot of CuPc microdisks were disrupted. Due to this reason, we replaced the use of ultrasound by an application of high-flow flush by acetone from a wash bottle. An influence of the resist thickness on the lift-off process was not observed. Moreover, it might be useful to mention that an easier lift-off was observed for silicon substrates than for the fused silica ones. In the latter case, a CuPc excess on top of the resist was more prone to survive on the sites near the microdisks.

**2.2 Sample characterization**. The following characterization techniques and instruments were utilized to control properties of PMMA resist films, CuPc films and CuPc microstructures during the individual fabrication steps. The resist thickness was measured by optical spectroscopic reflectometry (NanoCalc 2000, Ocean Optics Inc.). Mechanical profilometry (Dektak XT, Bruker Corporation) and scanning probe microscopy (Dimension Icon, Bruker Corporation) were used for measuring the CuPc film thickness and disk heights. Scanning probe microscope was also used for analysis of CuPc surface properties together with SEM in a dual-beam FIB-SEM Lyra3 setup (Tescan Orsay Holding), and optical microscopy. In addition, the information about the crystalline structure of CuPc thin films was obtained using by micro-Raman spectroscopy (alpha300 R, WITec Wissenschaftliche Instrumente und Technologie GmbH) using a green laser with a wavelength of 532 nm, 100× objective (Zeiss, NA 0.9, WD 0.31 mm), UHTS visual spectrometer (used gratings of 600 or 1800 grooves/mm), and back illuminated EMCCD camera. The typical laser power was set to the low value of 0.3 mW. In the case of the local warming experiment, the laser power was tuned up to 30 mW.

## 3. RESULTS AND DISCUSSION

An ordered 3×3 mm$^2$ square array of CuPc microdisks (diameter 5 μm, pitch 10 μm, Fig. 2) was fabricated on the silicon or fused silica substrates using EBL, CuPc deposition and subsequent resist removal by lift-off technique. According to the deposition parameters (deposition rate, duration, and temperature of the crucible with CuPc powder), films thickness and comparable disk heights were achieved from 7 up to 70 nm. The fabricated microdisks or films with thickness of 70 nm have green-yellowish and pink colours under optical microscope when fabricated on fused silica or silicon substrates, respectively (see comparison of colours in Fig. S1). They appear very faint for thicknesses below 20 nm.

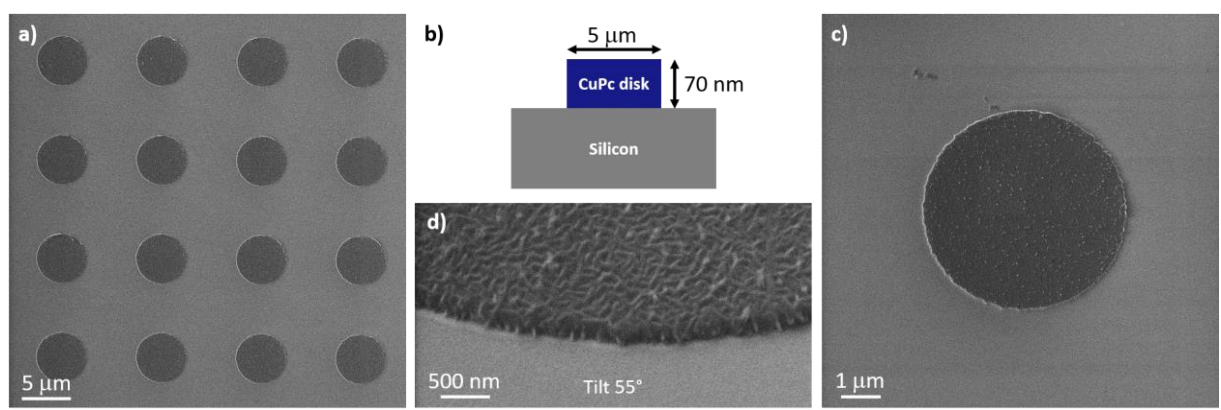

**Fig. 2.** SEM micrographs of CuPc microdisks fabricated on a single crystal silicon substrate. (a) Selected subarray of 4×4 microdisks. (b) Schematic of a single microdisk. (c) Detail view of a single microdisk. (d) Edge of the microdisk (side view, tilt of 55°).

Surface morphology of continuous CuPc films differs according to the film thickness (see Fig. 2d, 3a, b or 4c, d). Ultrathin films (7 nm, Fig. 3a, 15 nm, Fig. 4c) contain small grains and a very faint indication of fibrous texture, which is much more pronounced in the case of thicker film (70 nm, Fig. 3b, 4d). Microdisks fabricated during the same deposition as the continuous 70-nm thick film (Si substrate was covered with patterned PMMA in this case) contain the similar fibrous texture and more particle-like islands over the texture on the CuPc surface when compared to the continuous CuPc film (Fig. 3b, c). This can be attributed to a lift-off product (re-sedimented CuPc nanoparticles released from the CuPc excess). The height of particle-like islands above CuPc surface was typically below 50 nm (Fig. 4d), however, it can reach even 100 nm. In addition, several disks contained sharp edges with a total height up to 300 nm (comparable with the thickness of the PMMA resist) which are signs for locally non-ideal lift-off (corrupted rest of the material deposited on the walls of the resist microwells). If necessary, this minor imperfection can be suppressed by fabricating resist microwells with an undercut profile which can be achieved using a double-layer PMMA.

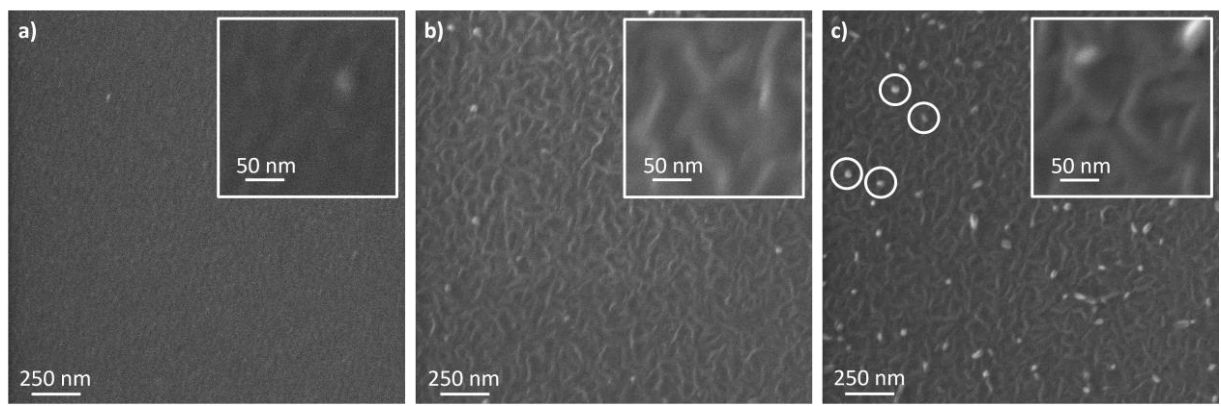

**Fig. 3.** SEM images of the CuPc surface on a single crystal silicon substrate, insets showing a detail of fibrous texture. (a) Continuous CuPc film with a thickness of 7 nm. (b) Continuous CuPc film with a thickness of 70 nm. (c) Surface of a final CuPc microdisk with a height of 70 nm done during the same deposition as the film in the case (b), the microdisk surface contains more particle-like islands (highlighted by the circles) above the fibrous texture than in the case of the continuous film.

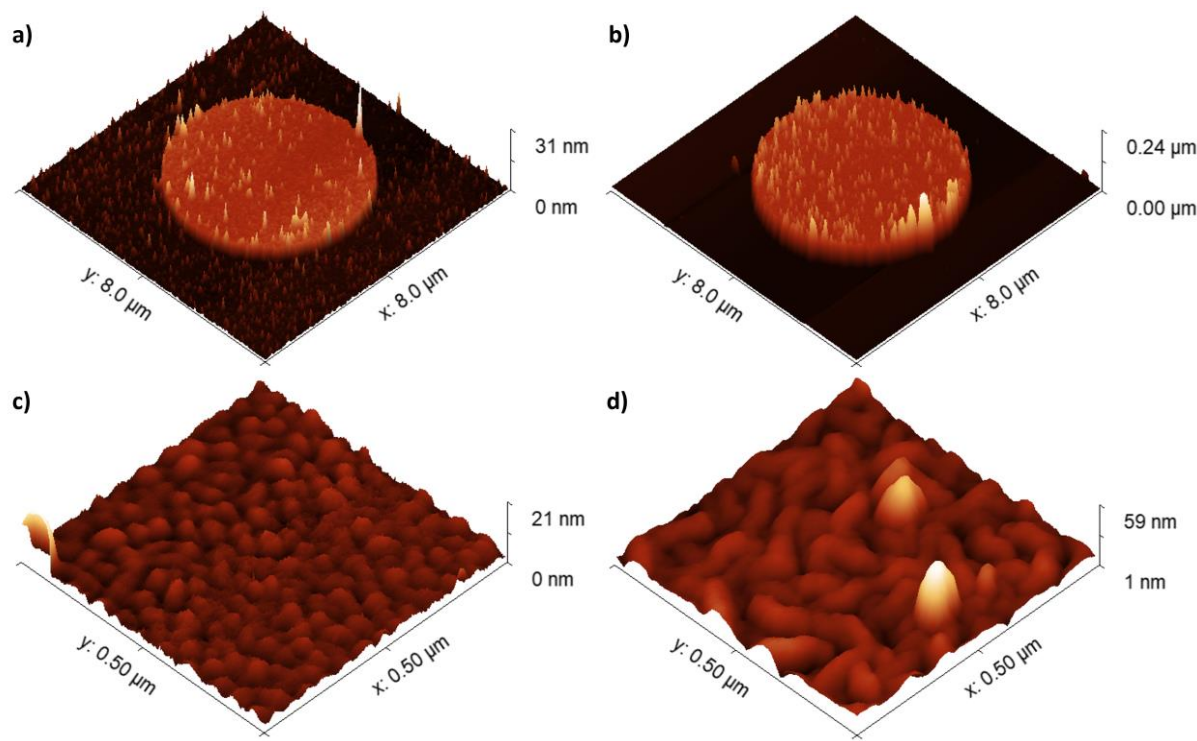

**Fig. 4.** AFM topography of CuPc disks and details of CuPc surface on a single crystal silicon substrate. (a) CuPc microdisk with a height of 7 nm. (b) CuPc microdisk with a height of 70 nm. (c) Detail of the continuous CuPc film with a thickness of 15 nm. (d) Detail of the CuPc microdisk with a height of 70 nm, there are two particle-like islands above the fibrous texture.

Despite the known low solubility of CuPc [51], the selection of the proper solvent (acetone) and therefore also of the appropriate resist (PMMA) was a critical part of the proposed fabrication procedure. Another positive tone resist CSAR 62 (AR-P 6200.09, Allresist GmbH) used for example in [52] and being recommended by its manufacturer for application with the AR 600-71 – dioxolane solvent (Allresist GmbH) was not usable, because it dissolves very quickly the CuPc film (even on the plain substrate without resist). The time scale for dissolving the CuP film (tens of seconds) was comparable with the time scale for the plain CSAR resist film.

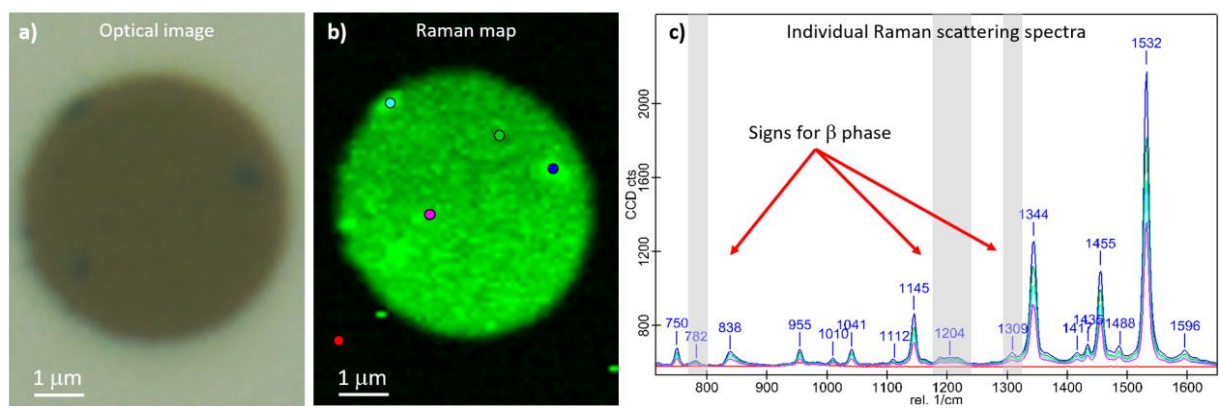

**Fig. 5.** Micro-Raman scattering spectroscopy of the CuPc disk with a height of 70 nm on the silicon substrate. (a) Optical image. (b) Micro-Raman map showing maximum intensity of the peak at 1532 rel. cm$^{-1}$. (c) Individual spectra obtained in the different areas of the same microdisk (colour of each spectrum corresponds to the same colour of the collection spot in the map), grey-colour highlighted peaks can be assigned to the β phase of CuPc.

To check the structural properties of these molecular microstructures, we have used micro-Raman scattering spectroscopy. The excitation laser radiation had a wavelength of 532 nm and the laser power was set to 0.3 mW to avoid degradation of the CuPc material. The observed point spectra from the surface of a CuPc microdisk (Fig. 5c) and CuPc film itself contained a series of Raman-active peaks known from literature [5,19,26,27,37,53–57]. According to Ghorai et al. the peaks at 782, 1204, and 1309 rel. cm$^{-1}$ ($E_g$, $B_{1g}$, and $B_{2g}$ modes) indicate the presence of the β phase of CuPc [19], whereas these peaks are missing in the case of the α phase. However, it should be mentioned that Anghelone et al. showed spectra containing these peaks even for the α phase, but the peaks were much weaker than in the case of the β and ε phase [5]. They proposed to use the ratio of areas under the selected pairs of peaks for phase classification. In our case, the ratios are on the limit between the β phase and presence of ε phase and the very weak peak at 300 rel. cm$^{-1}$ should exclude the ε phase [5] in our samples (Fig. 6a). The presence of a thermally stable β phase after deposition on the substrates kept during vacuum evaporation at room temperature (without post-annealing) was not reported yet; usually the metastable α phase is formed [38]. However, we used an extra low deposition pressure in the range of $10^{-6}$-$10^{-7}$ Pa and very low deposition rates both for thicker (1.0–1.7 nm/h = 0.017–0.028 nm/min) and thinner films (30 nm/h = 0.5 nm/min) allowing the molecules to arrange into an energetically preferable configuration. These rates especially for thicker films are more than one order of magnitude lower than those used to obtain films in the α phase, i.e. 0.8 nm/min reported in Ref. [6]. Nevertheless, other factors such as the type of substrate (crystalline or amorphous) could play a role as well.

The obtained micro-Raman map (Fig. 5b) confirms the presence of CuPc only in the area of the microdisk while the surrounding is free of CuPc. The CuPc is rather homogeneously distributed across the microdisk with the highest signal in the case of the darker areas apparent in the optical image (Fig. 5a) presumably related to the presence of large particle-like islands above the microdisk surface. The Raman signal is up to two times higher at these areas than at the remaining flatter microdisk areas (Fig. 5c) which is attributed to the larger local CuPc volume in these areas. This confirms that the particle-like islands with height up to 100 nm are formed also by CuPc. In addition, comparison of the micro-Raman spectra for CuPc films of various thicknesses on the silicon substrate shows no difference in peak positions – there is only much smaller signal from thin films (7 or 15 nm) than for the thicker ones (70 nm). In addition, no significant differences in the Raman spectra between the continuous CuPc film and the CuPc microdisks (samples without and with the application of acetone during the lift-off), or CuPc on both substrates (silicon, fused silica) have been observed (Fig. 6a, see the highlighted tiny differences).

To study the influence of local CuPc heating and to confirm the presence of the thermally stable CuPc phase, we carried out the following micro-Raman spectroscopic experiment using a variable laser power. There is an expectable upper power limit related to the local CuPc degradation (more about degradation below) and we varied the power in a range from 0.3 mW up to 30 mW to locally warm up a CuPc microdisk on the fused silica substrate. Increasing power of the laser radiation illuminating a small area of the disks, caused a significant shift in the position of the phonon $B_{1g}$ peak from 1531.7 to 1524.6 rel. cm$^{-1}$ ($\Delta\omega = 7.1$ cm$^{-1}$) and accompanied by the change of the peak width from 4.2 to 5.5 cm$^{-1}$. The results of this experiment are shown in Fig. 6b. Similar variations in the peak position ($\Delta\omega \sim$ 6 cm$^{-1}$) and peak width as a function of the sample temperature for the same Raman peak were obtained by Ghorai et al. [19]. However, in their case, a constant laser power of 0.05 mW was used and the whole CuPc sample was heated using an external heater. In contrast to our smooth power dependence, they observed a pronounced discontinuity in the peak position ($\Delta\omega \sim 1$ cm$^{-1}$) near a temperature of 183°C (456 K) which reflects the phase transition from the α to the β phase. Hence, the smooth dependence of the peak position in our experiment indicates the presence of the thermally stable phase (i.e. the β phase). Thus, this power-dependent Raman spectroscopy can be used as a complementary and reasonably-fast tool for excluding the occurrence of the thermally metastable α phase.

In addition, Ghorai et al. [19] mentioned that only the peak corresponding to 1532 rel. cm$^{-1}$ showed such a large thermal peak shift, while other peaks exhibited lower maximum shifts ($\Delta\omega$ 1–2 cm$^{-1}$). Significantly lower shifts in other peaks ($\Delta\omega$ up to 3.5 cm$^{-1}$) were observed also in our case with the increasing laser power (example in Fig. S2a). Only the thermally stable peak at 595 rel. cm$^{-1}$ showed an extra small shift ($\Delta\omega \sim 0.5$ cm$^{-1}$, Fig. S2b), which probably corresponds to the $A_{1g}$ vibration mode [27], a macro-cycle ring breathing mode [53].

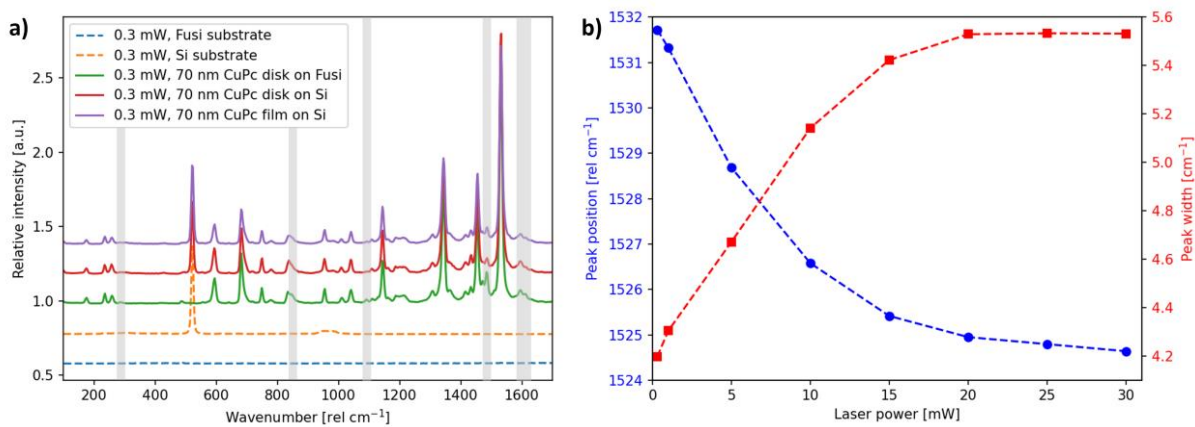

**Fig. 6.** Micro-Raman scattering spectroscopy. (a) Comparison of spectra for CuPc disks (fused silica- and silicon substrates) and 70 nm-thin film (silicon substrate). Grey-colour bands highlight minor differences between the CuPc disks on the fused silica- and silicon substrate. (b) Position (blue circles) and width (red squares) of the peak for the phonon $B_{1g}$ mode observed as a function of the laser power for a CuPc disk with a height of 70 nm on the fused silica substrate.

As demonstrated above, increasing laser power causes changes in the position and width of the peaks which is also accompanied by a growth of the peak intensities. This is happening only up to a certain value of power, after which the intensities stop growing and start to decrease, and the peak widths and positions return to the original values. This is well documented by measurements in which the laser power was first increased and then decreased (Fig. 7). For the lower maximum power (5 mW in Fig.7a), the peak position and width changes are reversible with respect to this increasing and decreasing phase and follow the same path. However, an application of higher maximum power caused irreversible changes as can be seen in Fig.7b for experiments with the maximum laser power 30 mW. This is true also for maximum intensity of the peak (Fig. 7c,d). The value of the maximum power triggering reversible changes is variable and mainly depends on the duration and cadence (switching strategy) of the laser irradiation, position at the CuPc microdisk (local molecular volume) and beam focus. In the case of combination of high laser power and continuing or high-cadence irradiation with a limited possibility to cooling down the heated microvolume, the absorbed energy causes local evaporation of CuPc molecules leading to a defect in the form of a crater (Fig. 7d, inset). Due to this evaporation, the irradiated CuPc volume decreases, thus the spectroscopic signal is significantly reduced (Fig. 7d), and peak changes stop to be reversible (Fig. 7b). However, the similar behaviour can be expected in the case that the whole CuPc film is warmed by external heater instead of local heating done by high-power laser.

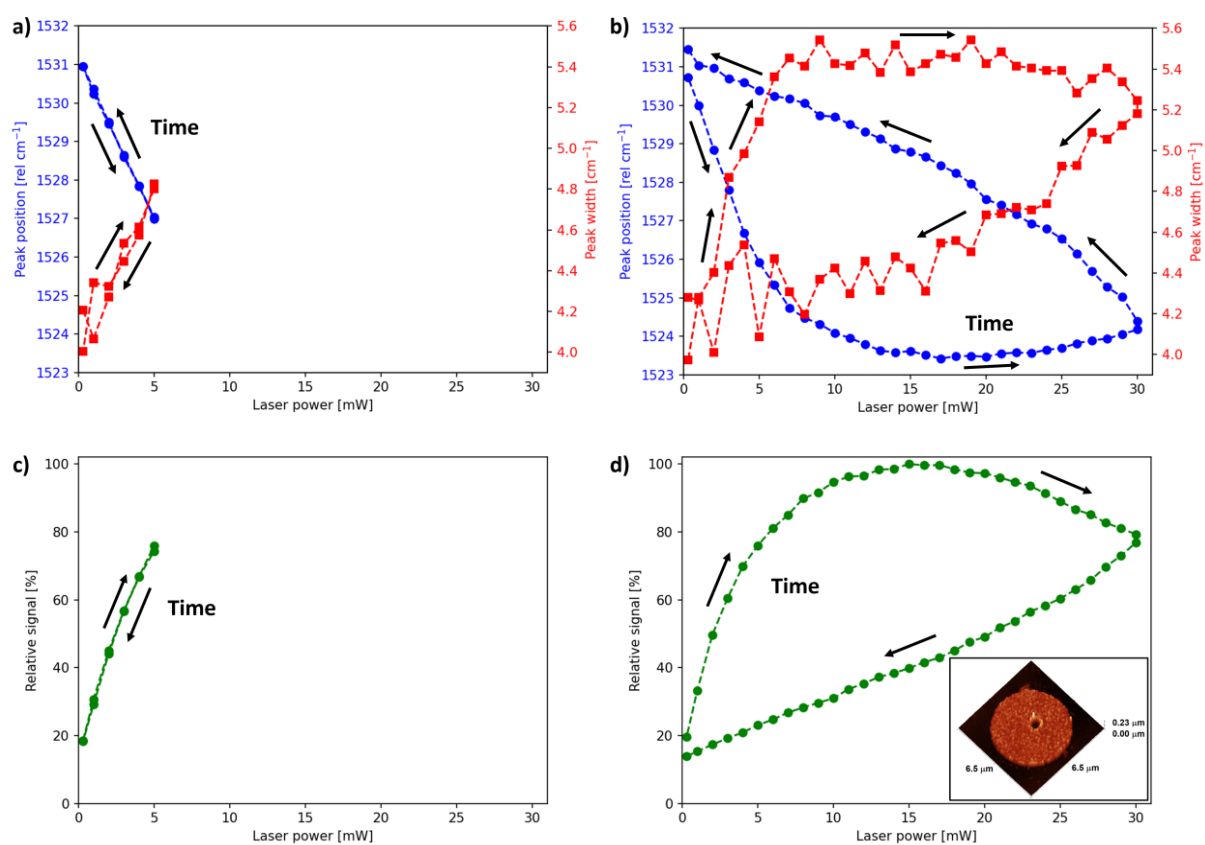

**Fig. 7.** Position (blue circles) and width (red squares) of the Raman peak (originally at 1532 rel. cm$^{-1}$) as a function of the laser power swept up to (a) 5 mW and (b) 30 mW. Variation of the maximum intensity of the same peak with the laser power swept up to (c) 5 mW and (d) 30 mW. Inset contains AFM picture of the microdisk with a crater after the laser experiment. CuPc disk with a height of 70 nm on the fused silica substrate.

## 4. CONCLUSIONS

To summarize our results, we have developed a procedure for the fabrication of arrays of CuPc microstructures (microdisks with a diameter of 5 μm, pitch of 10 μm and heights from 7 to 70 nm) by EBL. The critical points include finding the suitable deposition parameters, selection of a proper resist and its removal. The versatile fabrication process brings new possibilities for the fabrication of various micro/nanostructures such as micromagnets, plasmonic structures suitable for the study of strong coupling effects in mid-IR, etc.

CuPc films and fabricated microstructures show fibrous texture, well apparent in the thicker films (thickness of 70 nm). Micro-Raman scattering spectroscopy confirmed the presence of the β phase of CuPc molecules and its homogeneous distribution in the whole microdisks. Influence of the local heating of CuPc disks by laser was also studied. The most of CuPc Raman peaks showed significant and smooth changes in the peak width and position while increasing laser power. The highest amplitudes (peak shift $\Delta\omega \sim 7.1$ cm$^{-1}$, width from 4.2 to 5.5 cm$^{-1}$) was observed for the biggest Raman peak at 1532 rel. cm$^{-1}$ corresponding to the $B_{1g}$ phonon mode. Other CuPc peaks moved with an approximately half amplitude, only the peak at 595 rel. cm$^{-1}$ was thermally stable. The high laser power can significantly influence the properties of local area of CuPc disks. When the CuPc material begins to evaporate, the changes in the peak width and position stop to be reversible and the signal in the spectrum decreases. Our findings on smooth high-amplitude peak shifts can be used as a tool for excluding the α-to-β phase transition and as a confirmation of the thermally stable β phase suitable for thermometric and other applications.

**AUTHOR INFORMATION**

**ORCID**

Jiří Liška: 0000-0001-7080-1565

Tomáš Krajňák: 0000-0003-0920-3608

Peter Kepič: 0000-0002-9098-1900

Martin Konečný: 0000-0002-3628-3343

Martin Hrtoň: 0000-0002-3264-4025

Vlastimil Křápek: 0000-0002-4047-8653

Zdeněk Nováček: 0000-0002-5450-8710

Lorenzo Tesi: 0000-0003-4001-8363

Joris van Slageren: 0000-0003-4001-8363

Jan Čechal: 0000-0003-4745-8441

Tomáš Šikola: 0000-0003-4217-2276

**DECLARATION OF INTEREST**

The authors declare that they have no known competing financial interests or personal relationships that could have appeared to influence the work reported in this paper.

**ACKNOWLEDGMENTS**

This research has received funding from the European Union's Horizon 2020 research and innovation programme under grant agreement No 767227, H2020-Twinning project No.810626 – SINNCE, and the support by the Czech Science Foundation (Grant No. 20-28573S) and Baden-Württemberg Stiftung Competence Network Quantum Technology, project MOLTRIQUSENS. We acknowledge CzechNanoLab Research Infrastructure supported by Ministry of Education, Youth and Sports of the Czech Republic (LM2018110). We would like to express thanks to Dr. Alois Nebojsa for his help with measurements and other colleagues from our research group and staff from CF Nano.

**CREDIT OF STATEMENT**

All authors participated in the interpretation of results and preparation of the manuscript. JL: management of the experiment, fabrication (EBL, lift-off), TK: fabrication (deposition), PK: fabrication (EBL, lift-off), MK, LT, ZN: optical and structural characterization and other measurements, MH: theory support, VK, JVS, JČ, TŠ: conceiving and coordination of the research.

# SUPPLEMENTARY MATERIAL

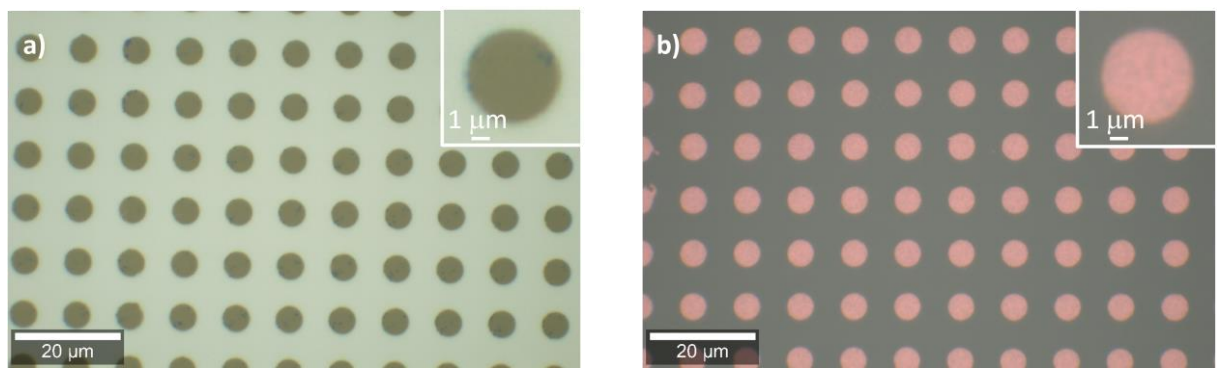

**Fig. S1.** Optical images of arrays of CuPc microdisks (diameter 5 μm, pitch 10 μm, height 70 nm). (a) On the silicon substrate. (b) On the fused silica substrate.

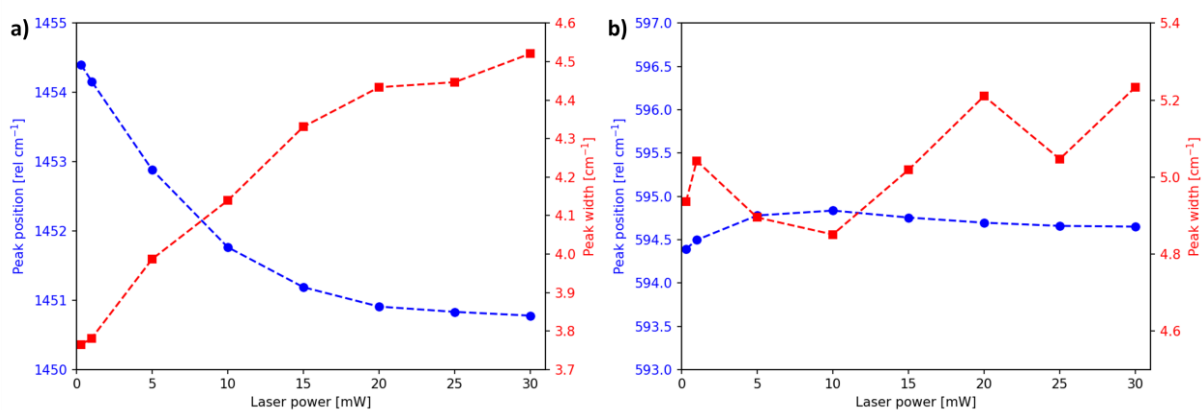

**Fig. S2.** Micro-Raman scattering spectroscopy. (a) Position and width of the peak for the phonon ($B_{1g}+B_{2g}$) mode observed using the variable laser power on the case of CuPc disk with a height of 70 nm on the fused silica substrate. (b) The same for the peak corresponding to the phonon mode $A_{1g}$.